\documentclass[9pt,twocolumn,twoside]{opticajnl}
\journal{opticajournal} % use for journal or Optica Open submissions

\setboolean{shortarticle}{true}
\usepackage{lineno}
\title{Spontaneous four-wave mixing in a thin layer with second-order nonlinearity}

\author[1,2,*]{Changjin Son}
\author[1,2]{Maria Chekhova}

\affil[1]{Max Planck Institute for the Science of Light, 91058 Erlangen, Germany}
\affil[2]{Friedrich-Alexander Universität Erlangen-Nürnberg, 91058 Erlangen, Germany}

\affil[*]{changjin.son@mpl.mpg.de}

\begin{abstract}
Pairs of entangled photons are crucial for photonic quantum technologies. 
The demand for integrability and multi-functionality suggests `flat’ platforms - ultrathin layers and metasurfaces - as sources of photon pairs. 
With the success in demonstrating spontaneous parametric down-conversion (SPDC) from such sources, an alternative process to generate photon pairs, spontaneous four-wave mixing (SFWM), also starts to attract interest.
In materials with nonzero second-order nonlinear susceptibility $\chi^{(2)}$, SFWM can  generate photon pairs both directly, through the third-order nonlinear susceptibility $\chi^{(3)}$, and in a cascaded way, through second harmonic generation (SHG) followed by SPDC. Usually, the cascaded process is more efficient. Here, we show that in a thin layer, direct SFWM dominates, because the wavevector mismatch for SFWM is much smaller than for SHG or SPDC. To demonstrate it, we implement the photon pair generation via SFWM in a second-order nonlinear material - a thin layer of lithium niobate (LN). The existence of both second- and third-order nonlinear processes offers broader opportunities for quantum state engineering. 

\end{abstract}

\setboolean{displaycopyright}{false} % Do not include copyright or licensing information in submission.

\begin{document}

\maketitle

\section{Introduction}
Spontaneous parametric down-conversion (SPDC) - a second-order nonlinear process where a single pump photon splits into a pair of daughter photons - is widely used to generate entangled photons in bulk crystals and waveguides. During the last few years, SPDC has been also implemented in micro- and nanoscale nonlinear layers \cite{okoth2019microscale,santiago2021entangled,Sultanov2022,guo2023ultrathin,sultanov2023temporally,weissflog2024tunable,Santos2024,feng2024polarization,Lyu2025,Liang2025,sorensen2025simple,Stich2026thin}.
Due to their tiny thickness, such layers satisfy the SPDC phase-matching automatically, even though the wavevector mismatch $\Delta k$ can be very large. As a result, materials with high second-order susceptibility $\chi^{(2)}$, or orientations using large components of the $\hat{\chi}^{(2)}$  tensor, which normally are excluded from SPDC because of phase-matching restriction, come into play. Although, because of the small thickness, such sources produce photon pairs at low rates -- up to a few kHz so far~\cite{Sultanov2024LC,Stich2026thin} -- their advantage is multifunctionality. For example, a thin layer can generate photon pairs not only forwards, but also backwards and bidirectionally~\cite{lu2025counter}; by tuning the pump polarization, the polarization state and the degree of entanglement of the photon pairs can be also tuned~\cite{Sultanov2022,weissflog2024tunable}; the pump wavelength and direction can be chosen arbitrarily.

%Utilizing an optical resonance is a breakthrough. Multiple works are demonstrated under the resonant situations, such as nano-resonators \cite{marino2019spontaneous} and metasurfaces \cite{santiago2021photon}. The optical resonances not only enhance the photon pair generation rate but also induce interesting functionalities: broad spatial entanglement \cite{zhang2022spatially}, emission direction control \cite{son2023photon}, and generation of complex entangled states \cite{santiago2022resonant}. Such research works have extended to another branch, which is to extremely thin layers, such as mono- \cite{lu2025counter} and a few atomic layers \cite{guo2023ultrathin}. A rising material is the so-called Van der Waals materials. The most famous one is transition metal chalcogenides (TMDs). Such materials' nonlinearity can be controlled by the stacking number of the atomic layers \cite{saynatjoki2017ultra}. Due to their inherent symmetry of the crystal structure, one can directly generate the Bell state \cite{weissflog2024tunable}.

The variety of materials used to generate entangled photons through SPDC is quickly expanding. Most promising are van-der-Waals crystals~\cite{guo2023ultrathin,weissflog2024tunable,feng2024polarization,Liang2025}, whose nonlinearity can be controlled by stacking several atomic layers in different ways~\cite{saynatjoki2017ultra}. Recently, entangled photons were generated in a liquid crystal~\cite{Sultanov2024LC,Klopcic2026LC}, which offers new ways to control the properties of entangled photons: through molecular orientation twist and by applying electric field.

%One inherent restriction is that $\chi^{(2)}$ materials should be non-centrosymmetric.
An alternative way to generate photon pairs is the spontaneous four-wave mixing (SFWM), where two pump photons are converted into a pair of entangled photons and the third-order susceptibility $\chi^{(3)}$ is involved. Since $\chi^{(3)}$ is nonzero in any material, the material choice becomes broader. An important advantage of SFWM is that it generates entangled photons on both sides of the pump spectral line. This allows one of the photons to cover blue or even UV range of wavelengths~\cite{Lopez-Huidobro2023UV_vis} without using even more blue-shifted pump.
During the last few years, considerable efforts to implement SFWM in `flat' sources have been made.
However, due to the weakness of the third-order nonlinear interaction compared to the second-order one, most of the experiments on SFWM in thin sources are carried out in the seeded regime~\cite{caspani2016enhanced,carnemolla2021visible,Xu2022FWM,Yang2025FWM,malek2025giant}, where a coherent beam at the frequency of one of the output photons is stimulating the process. %The situation is not much different under the resonances. The optical resonances enhance the FWM process \cite{nielsen2017giant,yang2022stimulated,malek2025giant}, but SFWM has not yet been demonstrated.
So far, there are only two experiments on photon pair generation via SFWM in thin layers~\cite{lee2017photon,son2025generation}.

Meanwhile, SFWM can also take place in non-centrosymmetric materials, where both $\chi^{(2)}$ and $\chi^{(3)}$ are present. Such materials were so far avoided in SFWM experiments. In the absence of the phase-matching restriction, a direct third-order nonlinear process should compete with a cascaded process.
For example, the third harmonic can be generated not only directly but also through the second harmonic generation (SHG) followed by sum frequency generation. This cascaded process is well-studied, both theoretically and experimentally~\cite{gennaro2022cascaded,saltiel2005multistep}.
Similarly, both cascaded and direct nonlinear processes can generate photon pairs at a certain wavelength set $\omega_s$ (signal) and $\omega_i$ (idler).
In the cascaded process, first, the pump at $\omega_p$ is up-converted to the second harmonic ($\omega_p + \omega_p \rightarrow \omega_{SH}$), and then the second harmonic (SH) photons are down-converted through SPDC: $\omega_{SH}\rightarrow \omega_s + \omega_i$.
The direct process generates photon pairs through SFWM: $2\omega_p\rightarrow \omega_s +\omega_i$.
The cascaded process is commonly considered to be stronger compared to the direct one.
However, its phase-matching condition is more restrictive than that of the direct process.

In this work, we show that in a thin nonlinear layer, the direct process is always stronger. To demonstrate this effect experimentally, we  use a $10$ $\mu$m-thick lithium niobate (LN) crystal and generate photon pairs through SFWM. 

%\textcolor{purple}{\it Changing the wavelengths won't help that much because the entire phase mismatch in the cascaded process has the same phase mismatch as that of the direct process. In my naive idea, it will be rather easier to use a chi-2 material where type-1 phase matching is possible for at least one of the sub-processes:$ee\rightarrow o, o\rightarrow ee$. While in such a source, SFWM will be done through $ee\rightarrow ee$. - CS}

\section{Experimental setup}
\begin{figure}[ht]
    \centering
    \includegraphics[width=0.8\linewidth]{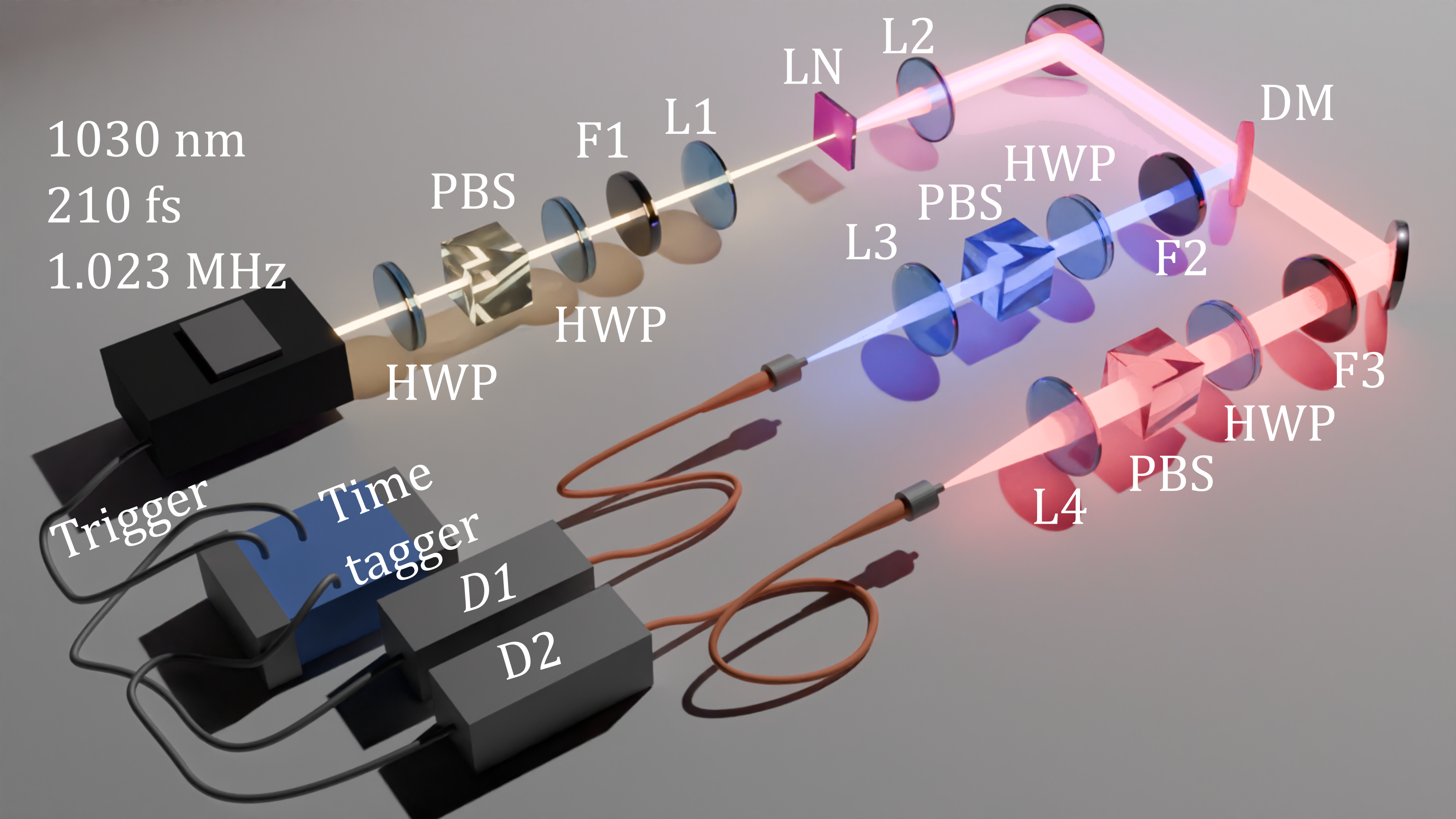}
    \caption{Experimental setup: lens L1 focuses the pump on the LN sample; signal (visible, shown in blue) and idler (IR, shown in orange) photons are collimated by lens L2, separated by dichroic mirror DM, and coupled into  detectors D1, D2 by lenses L3, L4, respectively. Their polarization is analyzed by an HWP and a PBS in each arm. Filter F1 eliminates unwanted wavelengths in the pump beam, and bandpass filters F2, F3 select the signal and idler photon detection bands. Photon detection events are registered by a time tagger triggered by pulses from the laser.}
    \label{fig: experimental setup}
\end{figure}

For the photon pair measurement, we use a Hanbury Brown\-- Twiss setup separating the entangled photon in two arms.
Figure~\ref{fig: experimental setup} shows the scheme of the experimental setup.
A laser at $1030$ nm, with a pulse duration of $210$ fs and a repetition rate of $1.023$ MHz, is used as the pump. Its
power and  polarization are  controlled by two half-wave plates (HWPs) and a polarizing beam splitter (PBS).
A band-pass filter (F1) at $1030$ nm with $10$ nm bandwidth filters out unwanted wavelengths.
The pump is focused by lens L1 into a $100\,\mu$m spot on the $10$ $\mu$m-thick x-cut LN layer on a $500$ $\mu$m-thick fused silica substrate, and the output photons are collimated by lens L2 with the focal length $100$ mm and NA=$0.5$.
The collected photons are separated by a long-pass dichroic mirror (DM) with the cut-off wavelength at $950$ nm.
The visible (reflected) and IR (transmitted) photons are coupled to the multi-mode fibers at the end of each arm by lenses L3 and L4, respectively.
At each arm, an HWP and a PBS are installed to analyze the polarization of the photons.
The detection wavelengths are selected by band-pass filters F2 ($770$ nm with $10$ nm band width) and F3 ($1550$ nm with $50$ nm band width), respectively.
The coupled photons are sent to single-photon detectors D1 (Perkin\&Elmer, SPCM-AQRH-16-FC) and D2 (IDQ, ID220), respectively. The detection events are triggered by the electric pulses of the laser and registered by the time tagger.

\section{Experimental results}
\begin{figure}[ht]
    \centering
    \includegraphics[width=\linewidth]{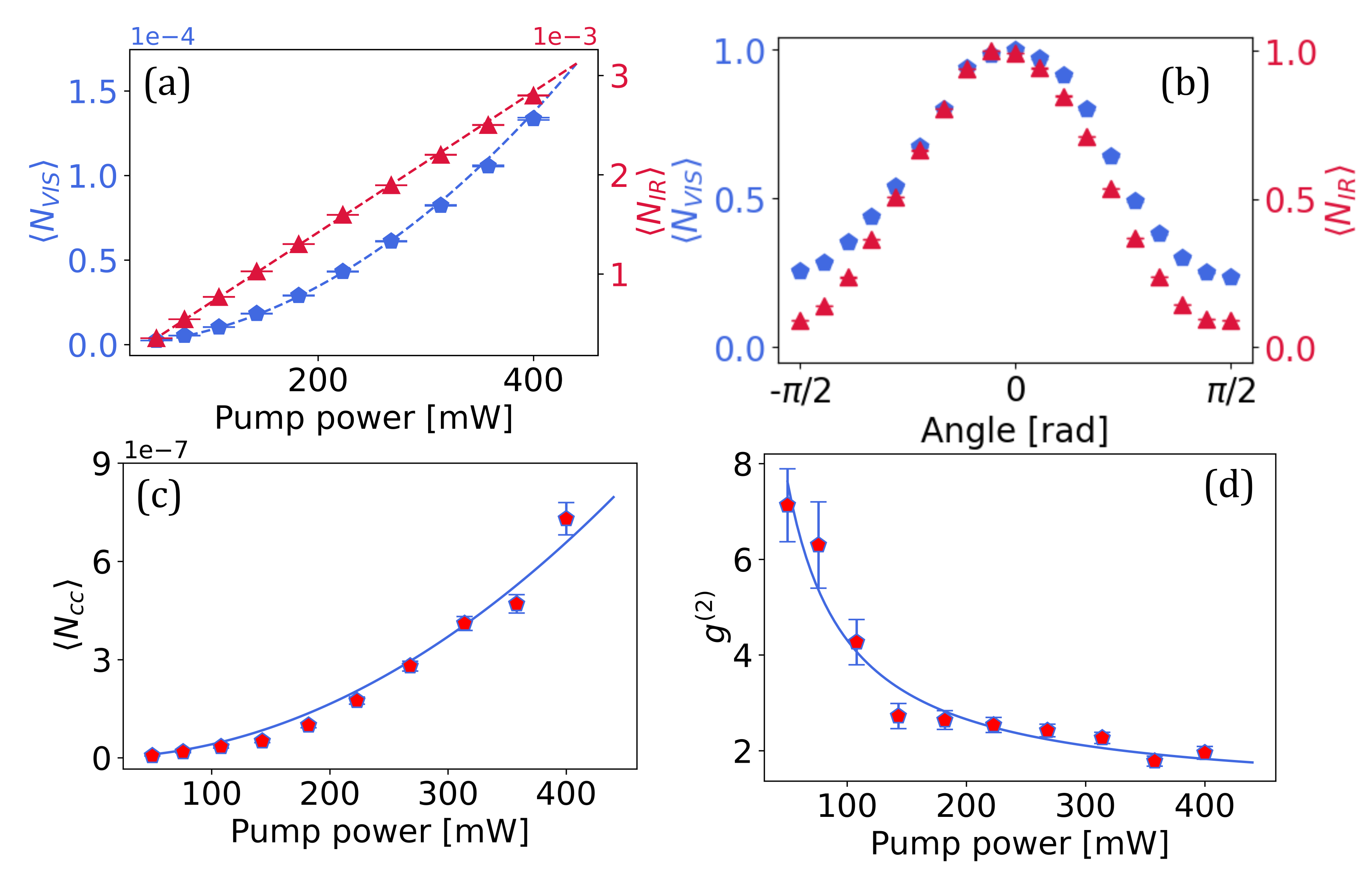}
    \caption{SFWM in LN. (a) Mean photon number per pulse for visible (blue circles) and IR (red triangles) photons. (b) The polarization dependences of the photon detection rates in the visible (blue circles) and IR (red triangles) arms. The $0$ angle corresponds to the pump polarization direction. (c) The number of coincidence counts per pulse as a function of the pump power, with a quadratic fitting curve. (d) Corresponding $g^{(2)}$ as a function of the pump power, with a fitting curve.}
    \label{fig: main results}
\end{figure}

Figure~\ref{fig: main results} shows the experimental results. The pump power dependences of the mean photon numbers per pulse for visible and IR photons, $\langle{N_{vis}}\rangle$ and $\langle{N_{IR}}\rangle$, are plotted in panel (a) with blue circles and red triangles, respectively.
The visible and IR photons exhibit different dependences on the pump power, one quadratic and the other linear.
The quadratic power dependence is indeed expected for SFWM. 
Meanwhile, the linear power scaling of the IR photons rate with the pump can be explained by SPDC. Indeed, under pumping at $1030$ nm, SPDC in thin LN can generate signal photons at $1550$ nm, with their idler counterparts being at $3.078\,\mu$m. The coherence length for this process is $15$ $\mu$m, longer than the sample thickness. While the idler photons cannot be detected by our setup, the signal photons generation rate should be much higher than the SFWM rate. Consequently, the dominant contribution to the rate of IR photons is from SPDC, which scales linearly with the pump power. Importantly, SPDC cannot affect the coincidence detection in our experiment because the idler SPDC photons are not detected in the visible arm.

Since the strongest components of LN second- and third-order nonlinear susceptibility tensors are $\chi^{(2)}_{zzz}$ and $\chi^{(3)}_{zzzz}$, respectively, both SPDC and SFWM are most efficient if the pump and the output photons are all polarized along the $z$ axis. Indeed, the photon count rates are maximal for the pump polarized along the $z$ axis (see Supplement 1). Also, from Fig.~\ref{fig: main results}(b), where the polarization dependences of $\langle{N_{vis}}\rangle$ and $\langle{N_{IR}}\rangle$ are plotted, we see that both visible and IR photons are polarized the same way as the pump. For visible photons, this is evidence that they mostly come from SFWM and not from two-photon photoluminescence.

Figure~\ref{fig: main results}(c) shows the mean number of coincidences per pulse $\langle{N_{cc}}\rangle$ as a function of the pump power. The dependence is quadratic, as expected for SFWM. The efficiency of pair generation is much lower than the one measured in SiN films using the same pump~\cite{son2025generation}; however, higher rates can be achieved because of the high damage threshold of LN at $1030$ nm. The second-order correlation function $g^{(2)}$, calculated as 
\begin{equation}
    g^{(2)}(0)=\frac{\langle{N_{cc}}\rangle}{\langle{N_{vis}}\rangle\langle{N_{IR}}\rangle},
    \label{eq:g2}
\end{equation}
is plotted in panel (d).  The fitting curve is $g^{(2)} = 1+a/P^2$, where $a$ is a fitting parameter and $P$ is the pump power. The $g^{(2)}$ exceeding the `thermal' value $2$ and its inverse dependence on the mean photon number of SFWM (scaling as $P^2$) indicate the detection of correlated photon pairs.

Importantly, the fused silica substrate could also contribute to SFWM, due to its noticeable $\chi^{(3)}=3\times10^{-22}$ m$^2$/V$^2$~\cite{boyd1992nonlinear} and large length. Its contribution can be estimated from these parameters and the phase matching condition (see the Supplementary Information) as $25$\% of the total rate of coincidences. 
%{Given that SFWM from fused silica does not depend on the pump polarization~\cite{son2025generation}, at least 50\%  of the observed effect comes from LN.}
 %The disagreement \textcolor{purple}{is} caused by the presence of SFWM mediated by the $\chi^{(3)}_{yyyy}$ component of LN, which is twice less than $\chi^{(3)}_{zzzz}$ \cite{schiek2025measurement}.

So far, it was not clear if the observed photon pairs are generated through the direct or cascaded process. To figure out the contribution of the cascaded process, we first investigated the efficiency of its second stage. To this end, using the same setup, we detected photon pairs from SPDC pumped at $515$ nm. The pump was prepared by frequency-doubling the $1030$ nm radiation in a bulk phase-matched BBO crystal installed after F1 (see Fig. \ref{fig: experimental setup}).
Further, knowing the efficiency of SHG from the LN layer, we estimated $\langle N_{cc}\rangle$ stemming from the cascaded process.
\begin{figure}[ht]
    \centering
    \includegraphics[width=0.94\linewidth]{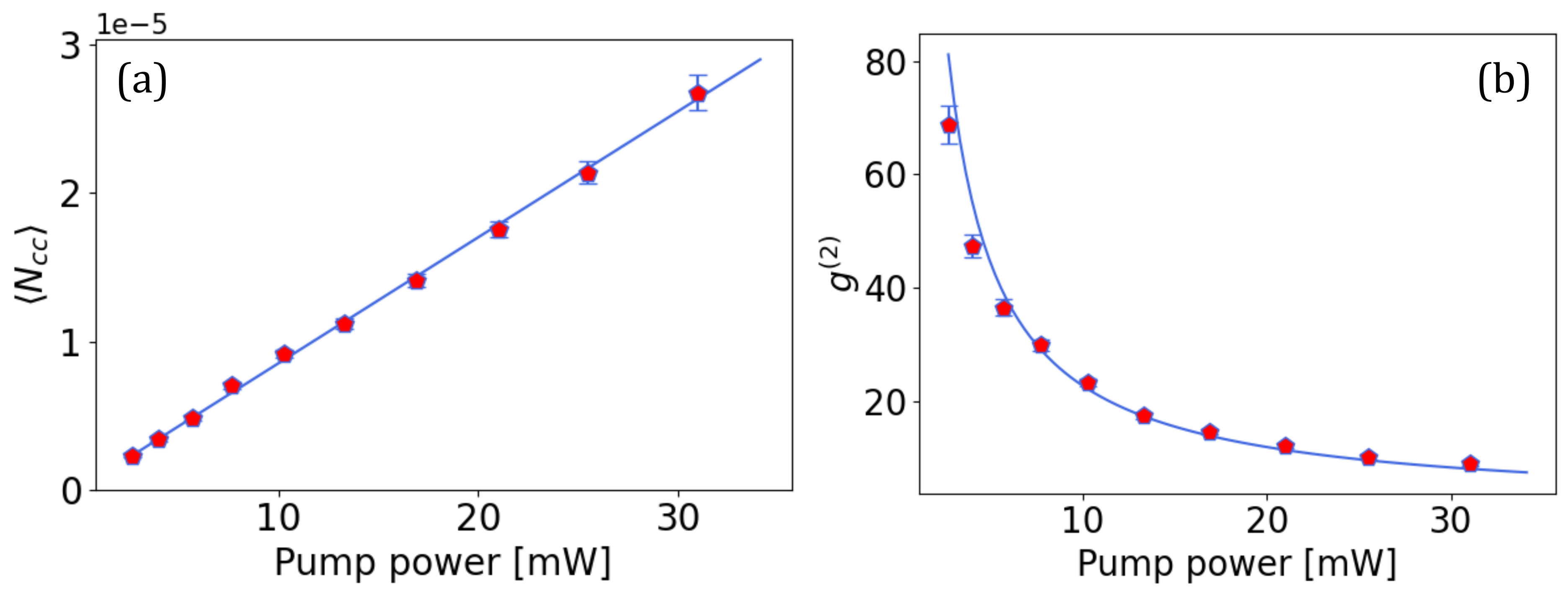}
    \caption{SPDC in LN pumped at $515$ nm. (a) The mean number of coincidence counts per pulse as a function of the pump power, with a linear fit. (b) Corresponding $g^{(2)}$ as a function of the pump power, with the fitting curve.}
    \label{fig: SPDC}
\end{figure}

Figure~\ref{fig: SPDC} shows the correlation measurement results for SPDC in the same LN layer, pumped at $515$ nm. The mean number of coincidences per pulse $\langle N_{cc}\rangle$ is plotted as a function of pump power with a linear fitting curve in panel (a). Panel (b) shows the correlation function $g^{(2)}$, calculated according to Eq.~\ref{eq:g2}, with the fitting curve  $g^{(2)}=1+a/P$. Meanwhile, the SHG efficiency of the LN layer is evaluated by measuring the pump and SH powers after the sample. The measured conversion efficiency is $2\times10^{-2}\%$/W: pumping with $400$ mW at $1030$ nm yields ~$30$ $\mu$W of SH. Based on this measurement, under $400$ mW pumping the cascaded process generates $4\times10^{-8}$ pairs per pulse, which is only 5\% of the total rate of photon pairs.

\section{Analysis and discussion}
\begin{figure}[ht]
    \centering
    \includegraphics[width=\linewidth]{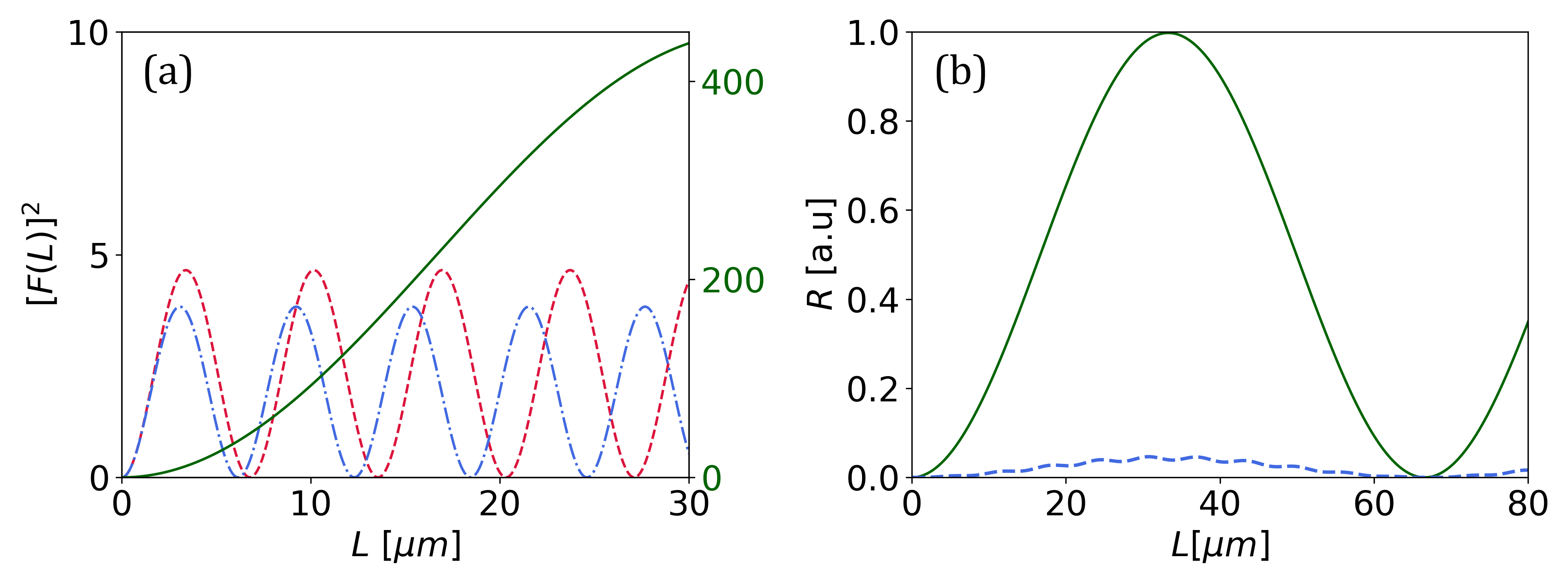}
    \caption{(a) $[F(L)]^2$ calculated for SFWM generating $770$ nm / $1550$ nm photon pairs from $1030$ nm pump (green solid line), SHG from $1030$ nm (blue dot-dashed line), and SPDC generating $770$ nm / $1550$ nm photon pairs from $515$ nm pump (red dashed line) as functions of the crystal length.
    (b) Calculated rates of photon pair generation through the direct (green) and cascaded (blue dashed line) processes vs the crystal length.}
    \label{fig: phase-matching}
\end{figure}

To describe the cascaded and direct processes theoretically, consider first their phase-matching functions 
\begin{equation}
F(L)=L\hbox{sinc}\frac{\Delta k L}{2},
    \label{PMF}
\end{equation}
where $\Delta k$ is the wavevector mismatch and $L$ the crystal length.  Figure~\ref{fig: phase-matching} shows $[F(L)]^2$ for each of the three involved processes: SFWM generating $770$ nm / $1550$ nm photon pairs from $1030$ nm pump (green solid line), SHG from $1030$ nm (blue dot-dashed line), and SPDC generating $770$ nm / $1550$ nm photon pairs from $515$ nm pump (red dashed line). For each process, the phase matching function oscillates (Maker fringes), reaching maxima at odd numbers of the nonlinear coherence length $L_{coh}=\pi/|\Delta k$|. The nonlinear coherence lengths for the three processes are $33.3$ $\mu$m, $3.1$ $\mu$m, and $3.4$ $\mu$m, respectively; therefore, the efficiencies of both steps of the cascaded process should oscillate faster than the efficiency of direct SFWM. Moreover, because the peak values of $F^2(L)$ scale inversely with $\Delta k^2$, they are much lower for SHG and SPDC than for SFWM. This already gives a qualitative explanation for the effect we see: because |$\Delta k_{SFWM}|=|\Delta k_{SHG}|-|\Delta k_{SPDC}|\ll|\Delta k_{SHG}|,|\Delta k_{SPDC}|$, the phase-matching function of SFWM considerably exceeds the ones of SHG and SPDC.  

For an accurate comparison, we consider both second-order processes, SHG and SPDC, occurring simultaneously. 
Assuming a non-depleted pump, the electric field amplitude of the SH generated from the pump propagating along the $x$-direction is \cite{boyd1992nonlinear}
\begin{equation}
    E_{SH}(x) =  \frac{2\pi}{\lambda_{SH}} \chi_{SHG}^{(2)}E_p^2\frac{e^{i\Delta k_{SHG}\cdot x}-1}{\Delta k_{SHG}},
    \label{eq. SHG}
\end{equation}
where $\lambda_{SH}$, $E_p$, and  $\chi_{SHG}^{(2)}$ are the SH wavelength, the amplitude of pump field, and the effective second-order nonlinear susceptibility for SHG, respectively.
The Hamiltonian of SPDC generating photons at frequencies $\omega_s$ and $\omega_i$ from the SH is \cite{rubin1994theory}
\begin{equation}
    \hat{H}_{SPDC}= \frac{2}{3} C(\omega_s,\omega_i)\int_{0}^{L} dx \cdot \chi_{SPDC}^{(2)}E_{SH}(x) e^{i\Delta k_{SPDC}\cdot x}a^{\dagger}_s a^{\dagger}_i+H.c.,
    \label{eq. SPDC}
\end{equation}
where $a^{\dagger}_{s,i}$ are the creation operators for the signal and idler photons, H.c. means Hermitian conjugation, $\chi^{(2)}_{SPDC}$ is the effective second-order nonlinear susceptibility for SPDC, and we have introduced $C(\omega_s,\omega_i)  \equiv  \hbar\sqrt{\omega_s\omega_i}/ (2\epsilon_0 V_q n_s n_i)$, with $n_{s,i}$ being the refractive indices at signal and idler wavelengths and $V_q$ the quantization volume. By substituting Eq.~(\ref{eq. SHG}) into Eq.~(\ref{eq. SPDC}) and integrating, we get the Hamiltonian of the cascaded process in the form 
\begin{equation}
 \hat{H}_{cas}= \frac{2}{3}C(\omega_s,\omega_i)  A_{cas}a_s^{\dagger}a_i^{\dagger}+H.c.,
        \label{eq: Hamiltonian SPDC}    
\end{equation}
with the amplitude 
\begin{eqnarray}
    A_{cas}= \frac{2\pi[\chi^{(2)}]^2E_p^2}{n_{SH}\lambda_{SH}\Delta k_{SHG}}  e^{i\Delta k_{SPDC}L/2}\nonumber
     \\
    \times \big[e^{i\Delta k_{SHG}L/2} F_{SFWM}(L)
     -F_{SPDC}(L)  \big],
        \label{eq: amplitude cascade}
\end{eqnarray}
where $n_{SH}$ is the refractive index at $\lambda_{SH}$, and we assumed $\chi^{(2)}_{SPDC}=\chi^{(2)}_{SHG}\equiv\chi^{(2)}$ and used the relation $\Delta k_{SFWM} = \Delta k_{SHG} + \Delta k_{SPDC}$. 

Meanwhile, the Hamiltonian for direct SFWM is  \cite{wang2001generation,PhysRevA.82.043809}
\begin{eqnarray}
    \hat{H}_{dir}= \frac{3}{4} C(\omega_s,\omega_i) \int_{0}^{L} dx \chi_{SFWM}^{(3)}E^2_p e^{i\Delta k_{SFWM} x}a^{\dagger}_s a^{\dagger}_i+H.c\nonumber\\
    = \frac{3}{4}C(\omega_s,\omega_i) A_{dir}a^{\dagger}_s a^{\dagger}_i+H.c,
    \label{eq:Ham dir}
\end{eqnarray}
with the amplitude
\begin{equation}
    A_{dir}=\chi_{SFWM}^{(3)} E^2_p  e^{i\Delta k_{SFWM}L/2}F_{SFWM}(L).
\end{equation}
Omitting the common factors in Eqs.~(\ref{eq: Hamiltonian SPDC},\ref{eq:Ham dir}), we see that the rates of pair generation through the cascaded and direct processes are, respectively,   $R_{cas}\propto\frac{4}{9}|A_{cas}|^2$ and $R_{dir}\propto\frac{9}{16}|A_{dir}|^2$.

Figure~\ref{fig: phase-matching} (b) shows these rates as functions of the crystal length: in green solid line for the direct process and in blue dot-dashed line for the cascaded process. For the calculation, we used  the values of $\chi^{(2)}=2.5\times10^{-11}$ m/V~\cite{shoji1997absolute} and $\chi_{SFWM}^{(3)}=1.5\times10^{-20}\,\mathrm{m^2/V^2}$~\cite{wang2019enhanced}. The rates oscillate with the same period, equal to twice the coherence length of the SFWM process. This is because the first term in the square brackets of Eq.~(\ref{eq: amplitude cascade}) dominates, see Fig.~\ref{fig: phase-matching}(b). At $10$ $\mu$m thickness, the ratio $R_{cas} /R_{dir}$ is $0.048$, which is in good correspondence to the experimental result.
%\textcolor{purple}{Meanwhile, the photon pairs can also be generated through SFWM from the fused silica substrate, which has a two orders of magnitude smaller $\chi^{(3)}$ value, $3.0\times10^{-22}\,\mathrm{m^2/V^2}$, however, has enough thickness, $500\, \mu\mathrm{m}$, to compensate it.
%From our calculation, $|A_{dir}|^2$ of LN is larger than that of the fused silica substrate by a factor of $2.4$. This value also corresponds to our rough estimation made through the polarization test of visible photons.
%By reflecting this fact, the estimated photon pair generation of the cascaded process to the total photon pair generation in LN becomes $7.1~\%$, not $5~\%$.}

%We showed that in a $\chi ^{(2)}$ material, the photon pairs can be generated by both cascaded and direct processes. 
If the cascaded process has the phase matching satisfied for both its stages (i.e., through periodic poling), it should be more efficient than direct SFWM. Note that for $\Delta k_{SH}=\Delta k_{SPDC}=0$, the field in Eq.~(\ref{eq. SHG}) scales as $z$, which leads to Eq.~(\ref{eq: amplitude cascade}) scaling as $L^2$ instead of $L$. But here we see that without phase matching satisfied, SHG and SPDC have larger wavevector mismatches than SPDC, which makes the cascaded process less efficient than the direct one.
%Although, entire phase-mismatch is the same as that of the direct process, the restriction at each sub-step affects the conversion efficiency.
%On the other hand, SFWM has a longer coherence length without such a restriction. 
%In a nutshell, the photon pairs generated through the cascaded and direct processes are spectrally and spatially indistinguishable; however, one can control the dominant process by carefully selecting the thickness of the nonlinear crystal.

SFWM in thin second-order nonlinear films can considerably enrich the toolbox of 'flat' quantum optics. In high-$\chi^{(2)}$ semiconductors, such as GaAs or AlGaAs, it can produce photon pairs with one of the photons above the bandgap, which is impossible through SPDC. Moreover, as we show above, a second-order nonlinear film pumped at frequency $\omega_p$ can simultaneously generate photon pairs at frequencies $\omega_s$ and $\omega_i$ through SFWM, but also at frequencies $\omega_i$ and $\omega_p-\omega_i$ through SPDC. This simultaneous action of two coherently pumped nonlinear effects can be used for building interesting quantum superpositions.

%\textcolor{blue}{\it The efficiency is lower than in SiN but the absolute rate can be higher because it can be pumped stronger.}

%\textcolor{purple}{Idea-Tuning the pump polarization and having polarization filters for output might be able to prepare a Bell state.}

\section{Back matter}

\begin{backmatter}
\bmsection{Funding}
Deutsche Forschungsgemeinschaft (568143457); ERC (Project 101199215 — MultiFlaQS).
\bmsection{Acknowledgment} 
We thank Francesco Tani, who kindly allowed us to
use his laser.

\bmsection{Disclosures} The authors declare no conflicts of interest.

\bmsection{Data Availability} 
Data underlying the results presented in this paper are not publicly available at this time but may be obtained from the authors upon reasonable request.

\bmsection{Supplemental document}
See Supplement 1 for supporting content.
\end{backmatter}

\section{References}
% Bibliography
\bibliography{References}

% Full bibliography added automatically for Optics Letters submissions; the following line will simply be ignored if submitting to other journals.
% Note that this extra page will not count against page length
\bibliographyfullrefs{References}

%Manual citation list
%\begin{thebibliography}{1}
%\bibitem{Zhang:14}
%Y.~Zhang, S.~Qiao, L.~Sun, Q.~W. Shi, W.~Huang, %L.~Li, and Z.~Yang,
 % \enquote{Photoinduced active terahertz metamaterials with nanostructured
  %vanadium dioxide film deposited by sol-gel method,} Opt. Express \textbf{22},
  %11070--11078 (2014).
%\end{thebibliography}

% Please include bios and photos of all authors for aop articles
\ifthenelse{\equal{\journalref}{aop}}{%
\section*{Author Biographies}
\begingroup
\setlength\intextsep{0pt}
\begin{minipage}[t][6.3cm][t]{1.0\textwidth} % Adjust height [6.3cm] as required for separation of bio photos.
  \begin{wrapfigure}{L}{0.25\textwidth}
    \includegraphics[width=0.25\textwidth]{john_smith.eps}
  \end{wrapfigure}
  \noindent
  {\bfseries John Smith} received his BSc (Mathematics) in 2000 from The University of Maryland. His research interests include lasers and optics.
\end{minipage}
\begin{minipage}{1.0\textwidth}
  \begin{wrapfigure}{L}{0.25\textwidth}
    \includegraphics[width=0.25\textwidth]{alice_smith.eps}
  \end{wrapfigure}
  \noindent
  {\bfseries Alice Smith} also received her BSc (Mathematics) in 2000 from The University of Maryland. Her research interests also include lasers and optics.
\end{minipage}
\endgroup
}{}

\end{document}